\documentclass[acmsmall,review,screen]{acmart}
\renewcommand\footnotetextcopyrightpermission[1]{}
\setcopyright{none}
\acmDOI{}
\acmISBN{}
\acmYear{2026}
\copyrightyear{2026}
\usepackage{tabularx}
\usepackage{subcaption}
\usepackage{array}
\usepackage{enumitem}
\newcolumntype{P}[1]{>{\raggedright\arraybackslash}p{#1}}
\newcolumntype{Y}{>{\raggedright\arraybackslash}X}

\author{Chijin Zhou}
\affiliation{
  \department{Shanghai Key Laboratory of Trustworthy Computing, Software Engineering Institute}
  \institution{East China Normal University}
  \city{Shanghai}
  \country{China}
}
\author{Quan Zhang}
\affiliation{
  \department{Shanghai Key Laboratory of Trustworthy Computing, Software Engineering Institute}
  \institution{East China Normal University}
  \city{Shanghai}
  \country{China}
}
\authornote{Quan Zhang is the corresponding author.}

\begin{document}
\title{Rethinking Access Control for AI Agents: Recent Advances and Open Challenges}
\authorsaddresses{}
\begin{abstract}
Large language model agents access data, invoke tools, and act on users' behalf. Keeping these actions within authorized bounds is challenging because agents choose their own execution paths while permissions and operating conditions may change. This paper surveys recent advances in agent access control across four areas: policy specification and derivation, identity and delegation, action and resource enforcement, and information-flow control. We compare their control objects, enforcement boundaries, and assumptions to examine what individual mechanisms achieve and how they can work together. Building on this analysis, we rethink four aspects of authorization: plan selection shapes permission requirements; contextual information can inform decisions without itself granting authority; restrictions must retain their meaning across system interfaces; and authorization must be reassessed as workflows evolve. These insights motivate open challenges and future directions for mechanism design and benchmark construction. In particular, evaluation should examine both observed agent behavior and the unauthorized effects that remain enabled. Our analysis connects existing mechanisms to the requirements for preserving authorization throughout an agent workflow.
\end{abstract}
\maketitle
\pagestyle{plain}
\thispagestyle{plain}
\raggedbottom
\section{Introduction}
Large language model (LLM) agents use tools to retrieve information, invoke services, and modify external state on a user's behalf~\cite{react,toolemu}. These capabilities allow agents to carry out tasks in diverse software environments~\cite{agentbench}, but also give their decisions consequences for protected data and resources. Access control determines which resources an agent may access, which operations it may perform, and under what conditions it may act~\cite{progent,saga}. The central requirement is to provide sufficient authority for the task while keeping execution within delegated bounds.

Authorizing an agent is difficult because both its course of action and the conditions governing permission can change during execution. Users typically specify a goal, while the agent selects concrete operations as it gathers information. Different paths to the same goal may access different resources, invoke different services, and produce different external effects; permission for one path need not cover another. New observations can lead the agent to revise its plan, delegation introduces additional principals, and delayed operations may outlive the approvals under which they were prepared. Authorization must therefore remain tied to the operations being performed and the conditions under which their effects occur. Successful task completion alone does not establish that the chosen execution stayed within the user's delegated authority.

Established security principles, such as least privilege~\cite{saltzer,privsep,capsicum} and complete mediation~\cite{saltzer,schneider,sasi}, have long guided the design of access control in software systems; applying these principles to agent workflows, however, requires accounting for decisions made by the agent itself. Least privilege limits authority to the permissions needed for execution, yet an agent can choose among plans with different resource and disclosure requirements. Complete mediation requires checking every protected operation, yet a single agent action may trigger effects across tools, delegated agents, and external services. Applying these principles therefore requires relating the agent's chosen plan to its authority requirements and tracing its actions to the boundaries where their effects can be controlled. In software engineering terms, this connects runtime enforcement to specifying evolving requirements~\cite{requirements} and governing third-party tool ecosystems~\cite{appstore}.

Recent research has advanced agent access control by addressing different aspects of authorization. Policy specification and derivation methods express constraints or infer them from task descriptions, organizational rules, and user preferences~\cite{conseca,agentspec,guardagent,shieldagent,miniscope,dataperms,drift}. Identity and delegation mechanisms establish who is acting, bind permissions to agents or tools, and constrain the scope of delegated access~\cite{delegation,saga,astra,aip,etdi}. Action and resource enforcement systems check operations at tool, application, browser, and operating-system boundaries~\cite{progent,ac4a,cellmate,agentbound,verisafe,csagent,pcas}. Information-flow control and supporting isolation mechanisms restrict how information reaches agent components, influences execution, or is released to other parties~\cite{camel,fides,ace,airgap,isolate,llmbda}. Together, these approaches provide complementary mechanisms for controlling agent authority throughout a workflow. Combining them requires preserving restrictions across interfaces: a policy limiting disclosure to an approved recipient, for example, must continue to constrain delegated agents and the tools through which they release information. Understanding these approaches requires examining what each mechanism enforces and how they compose.

In this paper, we aim to clarify the capabilities and limitations of existing approaches to agent access control and identify further directions. Our main contributions are as follows:
\begin{itemize}[leftmargin=*]
    \item \textbf{Surveying.} We survey and classify recent work on agent access control into the four mechanism families above. We compare their control objects, enforcement boundaries, and assumptions, and examine how these mechanisms jointly preserve authorization throughout an agent workflow.
    \item \textbf{Rethinking.} We draw on this analysis to rethink four aspects of agent access control: how plan selection shapes the authority an agent needs; when contextual information may inform authorization decisions; how restrictions survive translation across system interfaces; and whether authorization remains valid as workflows evolve.
    \item \textbf{Forecasting.} Building on these insights, we identify open challenges and directions for future research across planning, context handling, tool contracts, recovery, and benchmark design.
\end{itemize}

\section{Agent Access Control}\label{sec:background}
\subsection{An Agentic Workflow}
An agentic workflow turns a user's goal into actions whose details emerge during execution. An orchestrating agent gathers information, selects tools, and revises its plan as results arrive. It may delegate parts of the task to other agents, each with its own identity and credentials. The authority available to the workflow therefore depends on the participating principals and services, as well as on the actions the agent selects.

Consider an assistant asked to prepare a report from approved records and deliver the reviewed result to a designated recipient. The assistant retrieves the records, prepares a draft, delegates review, and sends the approved report. It can process the records locally or use a hosted service, but the latter transfers inputs to another service. Delegated review introduces another principal, and delivery releases information outside the organization. These choices make authorization relevant throughout the workflow, from selecting a processing strategy to releasing the result.

The workflow can also outlast the conditions under which an action was approved. A changed contact or withdrawn approval may invalidate a queued message, while a failed acknowledgement can leave delivery uncertain. The assistant must account for these changes alongside delegated work, acquired data, and pending effects.

\subsection{Authorization in Agentic Workflows}
\emph{Agent access control} determines which effects an agent workflow is authorized to produce. It governs access to resources, tool use, information propagation, and delegation. Principals include users, orchestrating and delegated agents, and services; organizational policies and resource owners impose additional constraints. Protected resources include source documents, derived drafts, credentials, and operations on external services. Tasks sharing one model can differ in identity and grants.

Within a workflow, \emph{authority} denotes the scope of resources a principal may access and effects it may produce, whereas a \emph{permission} is a policy unit representing a specific allowed operation or access. A task \emph{grant} (or \emph{authorization}) delegates part of the user's authority to an agent under stated conditions, bounded by organizational policy, while an \emph{approval} is an explicit confirmation from a user or designated reviewer that permits a sensitive operation or expands that grant. Tool availability shows what can be invoked, while authorization governs what is allowed.

\begin{figure}[t]
\centering
\Description{The illustrated agent workflow separates organization resources and execution, a delegated reviewer, and external services. Dashed arrows carry grants and policy updates; solid arrows show execution or data exchange. Numbered checks concern read scope, delegation, data flow to a hosted processor, and release to the authorized recipient.}
\includegraphics[width=\linewidth]{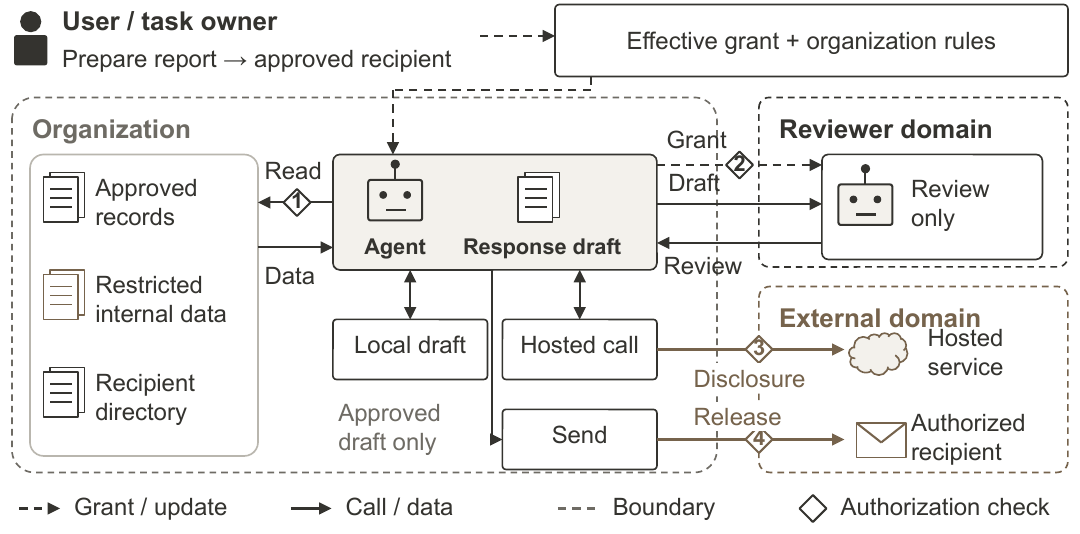}
\caption{Authorization boundaries in an illustrative agent workflow: (1) resource access, (2) scoped delegation, (3) hosted-service disclosure, and (4) external release.}
\label{fig:workflow}
\end{figure}

An agent workflow requires separate authorization decisions at each trust boundary. As Figure~\ref{fig:workflow} shows, reading a record requires source access, using a hosted processor requires permission to disclose its inputs, and delegating review requires authority to share the draft with another principal. Final delivery must satisfy the grant's content, recipient, and approval conditions. A valid permission at one boundary does not establish permission at the next: access to a document, for example, need not include authority to transmit it.

Enforcing the grant requires coordinated controls over principals, operations, and data. Authentication establishes who is acting; action policies constrain what that principal may do; information-flow policies govern how accessed data may be used and released. These controls operate at the interfaces where resources are read, tools execute, and information crosses trust boundaries~\cite{saltzer,denning}. Their responsibilities must be explicit so that restrictions remain enforceable across components.

\subsection{What Makes Agent Access Control Challenging?}
Unlike conventional distributed systems, where call graphs are fixed in code and control logic is separated from untrusted data, agent workflows synthesize control flow at runtime from natural-language context that mixes instructions with external observations. This execution model raises three fundamental challenges for agent access control.

First, an agent's \textbf{execution plan is often incomplete} when authority is delegated. It may discover a resource or select a different tool only after examining intermediate results. An initial grant must support useful discovery while bounding the actions that can follow. This connects agent authorization to the requirements problem identified by Harel and Marron: intended behavior may be incomplete, evolving, and difficult to articulate before interaction with the environment~\cite{requirements}.

Second, information obtained during execution can \textbf{influence decisions without granting authority}. A retrieved document may supply facts for the report, while a trusted directory may supply \emph{evidence} to resolve the authorized recipient. Neither source automatically holds an \emph{authorized role} that permits it to approve additional disclosure. Distinguishing these roles requires knowing where the information comes from (its provenance) and how each source is permitted to influence decisions. The challenge is to support legitimate adaptation while preventing contextual instructions from acquiring authority merely by entering the agent's context~\cite{conseca,camel,fides}.

Third, authorization must remain valid across \textbf{concurrent work and delayed effects}. A delegated agent can continue after the parent changes its plan, a cached result can survive loss of source access, and a retry can repeat an operation whose outcome is unknown. Checking the initiating call alone cannot resolve these dependencies. Access control must account for active delegates, derived data, and pending operations throughout execution. The next section examines mechanisms that address different parts of this authorization chain.

\section{Recent Advances}\label{sec:advances}
\begin{figure}[t]
\centering
\Description{An authorization chain links a task request, explicit policy, scoped identity, tool operation, and effect. Policy derivation, delegation, and action enforcement control related interfaces; information-flow control spans reading, transformation, and release. Feedback carries policy updates and revocation.}
\includegraphics[width=\linewidth]{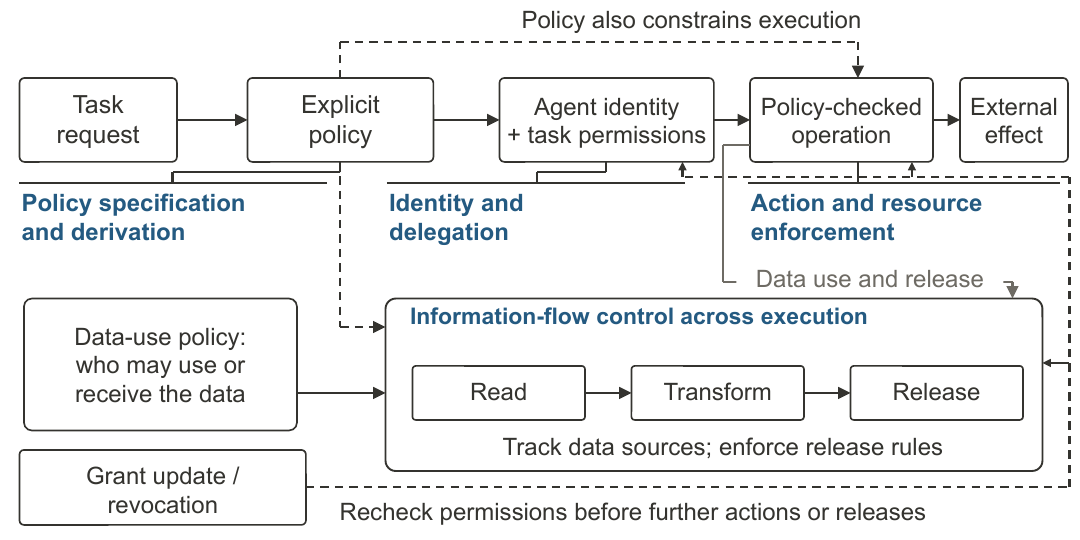}
\caption{Four complementary mechanisms for agent access control. Policy specification defines restrictions, identity and delegation assign task permissions, and action enforcement checks operations. Information-flow control governs data use and release throughout execution, while permission changes trigger renewed checks.}
\label{fig:mechanisms}
\end{figure}

We survey recent work on agent access control across 2024--2026 security, software engineering, and machine learning venues as well as recent preprints proposing concrete enforcement mechanisms, organizing the literature into four categories: policy specification and derivation, identity and delegation, action and resource enforcement, and information-flow control. Integrated systems may span multiple categories. As Figure~\ref{fig:mechanisms} illustrates, these four families trace the lifecycle of an agent workflow: policy specification defines task restrictions, identity and delegation bind those restrictions to acting principals, action enforcement mediates concrete tool and resource operations, and information-flow control governs how accessed data propagate and leave the system. Table~\ref{tab:mechanisms} summarizes representative approaches, where the last two columns specify the runtime artifacts and boundaries each mechanism directly mediates (\emph{mechanism and control object}) and the underlying assumptions required for its guarantees to hold (\emph{key dependency}).

\begin{table}[!tb]
\caption{Recent approaches grouped by authorization role, with their control boundaries and dependencies (publication venues and arXiv preprint versions are detailed in the bibliography).}
\label{tab:mechanisms}
\small
\setlength{\tabcolsep}{4pt}
\begin{tabularx}{\linewidth}{@{}P{.34\linewidth}P{.31\linewidth}Y@{}}
\toprule
Approach and works & Mechanism and control object & Key dependency \\
\midrule
\multicolumn{3}{@{}l}{\textbf{Policy Specification and Derivation}} \\
Explicit rules~\cite{agentspec,guardagent,shieldagent} & Rules, code, logical circuits & Rule fidelity; predicate accuracy \\
Task-derived constraints~\cite{conseca,drift,taskshield} & Context, trajectory, relevance & Trusted context; task authority \\
Plan-aware permissions~\cite{miniscope,prudentia} & Role selection; policy-aware plans & Role granularity; available plans \\
User preferences~\cite{dataperms,janus,userpolicies} & Predicted choices; approval rules & Preference coverage; user input \\
\midrule
\multicolumn{3}{@{}l}{\textbf{Identity and Delegation}} \\
Identity binding~\cite{delegation,saga,etdi} & Agent identities; tool signatures & Identity roots; grant interpretation \\
Scoped delegation~\cite{astra,aip,capmas} & Issuance; token attenuation & Scope derivation; token verification \\
Credential separation~\cite{miniscope,sudp} & Session proxy; secret custodian & Mediated use; operation binding \\
State and revocation~\cite{bounded,vera} & Sessions; delegation graphs & State coordination; revocation rules \\
\midrule
\multicolumn{3}{@{}l}{\textbf{Action and Resource Enforcement}} \\
Call checks~\cite{progent,toolfence} & Tool calls and arguments & Update authority; tool semantics \\
History-based checks~\cite{seagent,pcas,vigil} & Interaction graphs; event traces & Event coverage; policy/fact fidelity \\
Application semantics~\cite{ac4a,verisafe,csagent} & Resource, GUI, context checks & Application models; context extraction \\
Execution confinement~\cite{cellmate,agentbound,ztac} & HTTP, process, object mediation & Mediation coverage; trusted broker \\
Skill governance~\cite{skillguard,skillscope} & Capability checks; skill patches & Capability inference; replay coverage \\
\midrule
\multicolumn{3}{@{}l}{\textbf{Information-Flow Control}} \\
Minimization and isolation~\cite{airgap,isolate} & Conversation and app exchanges & Data selection; context isolation \\
Tracked computation~\cite{camel,fides,ace,llmbda} & Labeled values and interpreters & Flow semantics; release rules \\
Cross-boundary flows~\cite{gaap,spides,multicamel,safeflow} & Task, browser, inter-agent flows & Provenance; label accuracy \\
Controlled continuation~\cite{prudentia,appa} & Endorsement; confined recovery & Exception authority; isolation \\
\bottomrule
\end{tabularx}
\end{table}

\subsection{Policy Specification and Derivation}
Policy specification and derivation turn task requirements and user choices into enforceable restrictions. Existing approaches differ in the evidence they use to determine restrictions and how they translate that evidence into policies or permission decisions.

\emph{Explicit rules} express requirements as executable checks. AgentSpec~\cite{agentspec} provides event-triggered rules, GuardAgent~\cite{guardagent} generates checking code, and ShieldAgent~\cite{shieldagent} extracts and refines logical rules from policy documents. These systems connect natural-language requirements to runtime decisions, including access restrictions within broader safety policies. The available predicates and observations determine which requirements these checks can express. Their correctness also depends on whether the translation preserves the intended restriction.

\emph{Task-derived constraints} reduce the need to author every restriction manually. Conseca~\cite{conseca} derives policies from designated trusted context, whereas DRIFT~\cite{drift} constructs function trajectories and parameter constraints that can be revised during execution. Task Shield~\cite{taskshield} checks contribution to the user's task, providing a relevance criterion rather than an independent grant of authority. \emph{Plan-aware permissions} link the authority an agent receives to the execution plan it selects. MiniScope~\cite{miniscope} selects permissions covering a proposed plan; Prudentia~\cite{prudentia} guides planning with policies and data labels specifying handling restrictions to reduce unnecessary approval requests. These planning approaches differ in whether they assign permissions to a plan or let policies guide planning.

\emph{User preferences}, elicited directly or predicted from context, supply information that task descriptions alone may omit. Data-access permission prediction learns contextual sharing preferences~\cite{dataperms}. Janus~\cite{janus} applies persistent rules before passing calls not approved by those rules to a permission assistant that can consult the user. Consequence-based policies let users express reusable allow, ask, or deny decisions~\cite{userpolicies}. These approaches address which permission decisions can be automated using established or predicted preferences and which require further user input.

\subsection{Identity and Delegation}
Identity and delegation connect an agent's actions to the principal it represents and the authority it receives. Existing approaches differ in how they establish delegation relationships, limit the authority passed to agents, and maintain those limits as execution proceeds.

\emph{Identity binding} establishes who acts on whose behalf, while delegation bounds the authority passed to each agent. Authenticated delegation proposals link users, agents, and permission scopes through identity protocols~\cite{delegation}. SAGA~\cite{saga} distinguishes permission to establish contact with an agent from authorization of subsequent requests. For tools, ETDI~\cite{etdi} uses provider-signed, versioned definitions to establish tool identity and declare required permissions. These records support authorization decisions, while services enforce the associated permission limits. Emerging industry standards such as the Model Context Protocol (MCP) authorization specification (OAuth~2.1) and the Agent-to-Agent (A2A) protocol standardize endpoint authentication and coarse-grained scopes, whereas the mechanisms above bind authority to fine-grained operations.

\emph{Scoped delegation} uses credentials to carry explicit limits on delegated authority. El Helou et al.~\cite{astra} issue tokens with permission scopes selected by their semantic match to the task. AIP~\cite{aip} binds delegated authority to specific invocations through signed credentials, while CAPMAS~\cite{capmas} combines semantic scope selection with attenuating Macaroon tokens. Scope selection may depend on a model, while cryptographic attenuation prevents further delegation from expanding the granted permissions. \emph{Credential separation} keeps service credentials with an intermediary rather than the agent. MiniScope~\cite{miniscope} mediates session-scoped requests without exposing reusable credentials to the agent. SUDP~\cite{sudp} binds user approval to a standardized operation description. A custodian holds the credentials, validates the grant, and executes the matching operation.

\emph{State and revocation} mechanisms track execution history, shared limits, and withdrawn grants to determine whether delegated authority remains valid. Bounded Agents~\cite{bounded} maintains session history, scope, and budgets to reject prohibited action combinations; its composition result assumes adequate restrictions and admission decisions made one at a time. VERA~\cite{vera} defines edge revocation over a delegation graph, retaining access to shared agents while at least one independent authorization path remains valid. Delegation thus requires both a verifiable scope and rules for maintaining that scope as the workflow changes.

\subsection{Action and Resource Enforcement}
Action and resource enforcement applies authorization policies at the boundaries where tools operate and resources are accessed. Existing approaches differ in the operations they observe, the context available for authorization decisions, and the effects they can prevent.

\emph{Call checks} enforce restrictions on proposed tool invocations and their arguments. Progent~\cite{progent} evaluates policy conditions on tool calls and requires approval for policy expansion. ToolFence~\cite{toolfence} checks where sensitive argument values come from and whether calls stay within assigned budgets, using structured rules that specify permitted tools and argument sources. \emph{History-based checks} represent execution history explicitly to evaluate relationships across calls. SEAgent~\cite{seagent} matches patterns in graphs of agent interactions, while Palumbo et al.~\cite{pcas} evaluate Datalog policies over event graphs. VIGIL~\cite{vigil} checks requirements on event ordering and value flows across observed tool calls. These methods enforce cross-call dependencies when the monitor observes them.

\emph{Application semantics} allow enforcement mechanisms to interpret calls in terms of the resources and state changes they affect. AC4A~\cite{ac4a} maps application operations to the permissions they require. VeriSafe Agent~\cite{verisafe} checks anticipated GUI state transitions against logical specifications, while CSAgent~\cite{csagent} uses an operating-system service to evaluate policies that depend on user intent and execution context. These methods rely on application models or runtime observations to identify the resources and effects of an operation.

\emph{Execution confinement} restricts the requests, processes, and resource accesses through which tools operate. ceLLMate~\cite{cellmate} intercepts browser HTTP requests, while AgentBound~\cite{agentbound} confines Model Context Protocol (MCP) server processes. ZTAC~\cite{ztac} resolves the target object before making an authorization decision and checks at use that the operation reaches the authorized object. These systems cover execution paths that framework-level call checks may not observe. Their precision depends on how policy restrictions map to the requests and objects controlled at these boundaries.

\emph{Skill governance} controls the capabilities exercised by reusable packages of instructions and code. SkillGuard~\cite{skillguard} checks capability use at runtime against skill manifests and session grants. SkillScope~\cite{skillscope} models actions in skill instructions and code, then uses task replay to assess whether candidate actions exceed the permissions needed for the task. The former mediates capability use during execution, while the latter modifies inspected skills to constrain the identified behaviors. The enforcement boundary determines which operations are visible and which effects can be blocked.

\subsection{Information-Flow Control}
Information-flow control governs how data may influence computation and reach other parties after initial access. Existing approaches differ in how they limit data exposure, track restrictions through computation, and determine when information may be passed onward.

\emph{Minimization and isolation} limit the data available to agent components and the paths through which it can be exchanged. AirGapAgent~\cite{airgap} limits private data supplied to a conversation, while IsolateGPT~\cite{isolate} separates application contexts and mediates communication. Extending these controls to subsequent computation requires tracking restrictions on data after access.

\emph{Tracked computation} carries data restrictions through processing by tracking where values come from and which policies apply to them. CaMeL~\cite{camel} separates planning from data processing and uses an interpreter to track dependencies between values and check policies at tool calls. Fides~\cite{fides} tracks confidentiality and integrity restrictions throughout the agent loop, while ACE~\cite{ace} combines planning, execution isolation, and flow verification. LLMbda~\cite{llmbda} formally defines how labels propagate through conversations and generated code. Its machine-checked noninterference results constrain how changes in protected inputs affect observable outputs within its model. These approaches differ in the dependencies they track and the rules governing release. Their policies also determine when restrictions may be relaxed: endorsement permits lower-trust data to influence protected decisions, while declassification permits policy-authorized release of restricted data.

\emph{Cross-boundary flows} require data restrictions to remain enforceable across tasks, interfaces, and agents. GAAP~\cite{gaap} maintains disclosure history across tasks. SPIDES~\cite{spides} propagates confidentiality and integrity labels from web-page elements to checks on browser actions such as typing and clicking. multi-CaMeL~\cite{multicamel} passes trusted instructions and tool-derived data through separate channels between agents, keeping the data out of the receiving planner's instruction context. SafeFlow~\cite{safeflow} propagates risk labels through a graph linking tasks, messages, tool events, and agent states. It uses a model to annotate risks and assemble evidence from messages, retrieved content, and tool traces into structured records for rule-based checks. Coverage depends on whether sources and restrictions remain identifiable and accurately represented at each boundary.

\emph{Controlled continuation} seeks ways to complete a task when information-flow restrictions block the current execution path. Prudentia~\cite{prudentia} plans around policy constraints and distinguishes permission to trust data from approval to perform an action. APPA~\cite{appa} checks proposed tool calls before execution and uses isolated branches and policy-governed recovery to support continued work. These mechanisms allow execution to proceed through alternatives or exceptions authorized by policy while retaining restrictions on the data involved.

Across the four categories reviewed in this section, existing mechanisms specify, delegate, enforce, and propagate authorization restrictions. Their composition raises a central question: \textbf{How can authorization restrictions preserve their meaning and remain effective as an agent's workflow evolves?} This question will guide our rethinking in Section~\ref{sec:rethink}.

\section{Rethinking Agent Access Control}\label{sec:rethink}
\begin{figure}[t]
\centering
\Description{Three agent-authorization thought experiments contrast the authority required by alternative plans, resolution of choices within a grant versus expansion of that grant, and preservation versus loss of recipient and content restrictions during policy translation.}
\begin{subfigure}[t]{.3\linewidth}
\centering\includegraphics[width=.92\linewidth]{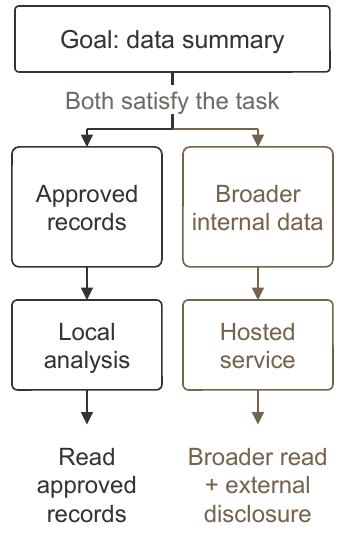}
\caption{Plan-dependent authority}\label{fig:plan}
\end{subfigure}\hfill
\begin{subfigure}[t]{.3\linewidth}
\centering\includegraphics[width=.92\linewidth]{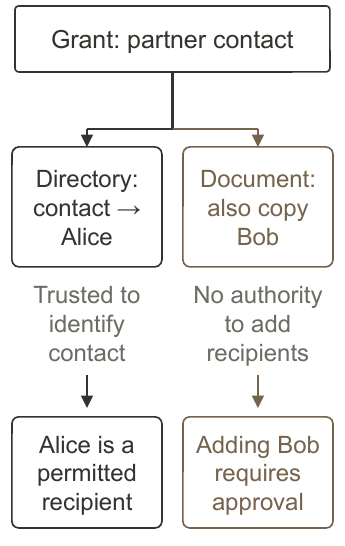}
\caption{Resolution versus expansion}\label{fig:context}
\end{subfigure}\hfill
\begin{subfigure}[t]{.3\linewidth}
\centering\includegraphics[width=.92\linewidth]{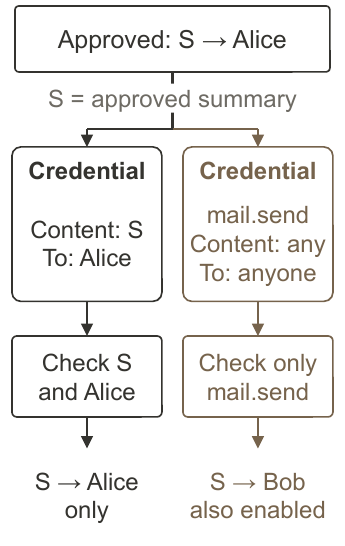}
\caption{Preserving policy meaning}\label{fig:representation}
\end{subfigure}
\caption{Three perspectives on preserving authorization: (a) plans that satisfy the same task can require different authority; (b) contextual evidence may identify an authorized recipient without authorizing additional recipients; and (c) recipient and content restrictions must survive translation into credentials and execution checks. $S$ denotes the approved summary.}
\label{fig:rethink}
\end{figure}

Preserving authorization throughout a workflow requires examining what authority is needed, how it is interpreted and enforced, and whether it remains valid as execution proceeds. Section~\ref{sec:rethink-plans} examines how plan selection determines the authority an agent needs. Section~\ref{sec:rethink-context} considers which contextual sources may resolve choices within a grant or authorize changes to it. Section~\ref{sec:rethink-representations} examines how approved restrictions retain their meaning across system interfaces. Section~\ref{sec:rethink-state} addresses how authorization remains valid across delegation, retries, and revocation. Together, these perspectives trace authorization from planning to ongoing execution, with concrete cases illustrated in Figures~\ref{fig:rethink} and~\ref{fig:state}.

% Keep the rethink figures after the section opening.
\suppressfloats[t]

\subsection{Least Privilege Begins with Plan Selection}\label{sec:rethink-plans}
A task's permission requirements depend on how the agent chooses to accomplish it; least privilege must therefore be considered during plan selection as well as permission allocation. Even an exact permission optimizer can allocate more authority than an adequate alternative plan would require if it starts from an unnecessarily intrusive strategy.

The report task illustrates how a plan can require permissions that an adequate alternative avoids. Figure~\ref{fig:rethink}(\subref{fig:plan}) compares two plans for the report task: one processes approved records locally; the other retrieves a broader dataset and sends it to a hosted analysis service. Suppose both meet the user's requirements for accuracy and completeness. Every permission assigned to the second plan might be indispensable to its execution, even though the first plan completes the task with less disclosure. Presenting only the second plan conflates its permission requirements with those of the task, excluding the less intrusive alternative from consideration.

Plan selection should therefore compare adequate alternatives against the current grant. Let $\Pi_q(g)$ denote candidate plans meeting a specified quality threshold $q$ for goal $g$, $A(\pi)$ the permissions required by plan $\pi$, and $G$ the permissions available under the currently valid grant. Planning under existing authority seeks
\begin{equation}
\pi \in \Pi_q(g) \quad\text{such that}\quad A(\pi)\subseteq G.
\label{eq:planning}
\end{equation}
Among feasible plans, selection should consider the access and disclosure each plan requires (authority exposure), execution cost, and user effort. If the search finds none, the system can propose a scoped grant change, negotiate a reduced outcome, or seek clarification.

Existing methods place permission reasoning at different stages of planning. MiniScope selects permissions for a proposed plan~\cite{miniscope}, Parkinson et al.\ propose exposing permission costs to the planner to support elastic least privilege~\cite{elasticprivilege}, and Prudentia uses policies and data labels to guide plan formation~\cite{prudentia}. This distinction matters because reducing a fixed plan's permissions and selecting a less intrusive plan within the current grant solve different problems. A planner can first seek alternatives satisfying Equation~\ref{eq:planning} and present any required expansion for separate approval.

Comparing plans requires an explicit account of both task quality and authority exposure. One plan may need sensitive local access, while another requires disclosure to an external service; neither is necessarily preferable under every policy. Where these costs are incomparable, a useful result is a set of adequate plans with explicit authority tradeoffs. Evaluation should measure both how closely granted permissions match a fixed plan's needs and the quality--authority tradeoff across plans; fixing the plan in advance measures only the former.

\subsection{Contextual Trust Must Be Tied to Decision Authority}\label{sec:rethink-context}
Contextual information can guide an agent's actions without carrying authority to change its permissions; trust must therefore be assessed in relation to the decisions a source is authorized to influence.
Resolving a permitted recipient and adding a new recipient require different authority. In Figure~\ref{fig:rethink}(\subref{fig:context}), a trusted directory identifies Alice as the partner contact permitted by the grant. A retrieved document asking the agent to also copy Bob instead proposes a different disclosure boundary. Both pieces of text are relevant to the task, but they play different roles: one supplies evidence for a permitted choice, while the other requests expanded authority.

Context-control mechanisms need an explicit account of each source's decision role. Conseca's trusted-context boundary, CaMeL's separation of planning and data processing, and Fides's selective inspection provide mechanisms for controlling contextual influence~\cite{conseca,camel,fides}. The permitted influence depends on the source's role: a document may contribute facts to a summary, a directory may establish a resource identifier, and a designated approver may authorize release. Preserving these roles allows useful contextual adaptation within explicit limits.

Restricting how a source expresses its response does not establish which decisions it may authorize. Suppose a workflow asks a source to answer only ``yes'' or ``no'' to ``is this draft approved for external release?'' The response carries little information but can determine whether confidential data leave the organization. Enforcing this format constrains the response, while authority to approve release must come from an applicable grant to the approver or an authorized decision procedure. Even a tightly constrained response therefore needs to be checked against the source's authorized role before it controls a consequential action.

These distinctions call for separate handling of decisions that resolve choices within an existing grant and decisions that expand that grant. Resolution requires evidence appropriate to the permitted choice; expansion requires an authorized source of change. Progent makes the latter boundary explicit by requiring approval for policy expansion~\cite{progent}. A workflow can also authorize adaptation in advance, for example by permitting instructions from a manager's task list within a bounded set of actions. In each case, the system needs to retain the provenance and authorized role of the evidence. Omitting source roles and approval links obscures the basis for later decisions.

\subsection{Preserving Authorization across Policy Translation and Execution}\label{sec:rethink-representations}
Individual components may enforce their local rules correctly even when restrictions are lost between them; preserving authorization therefore requires examining how an approved policy governs the resources accessed and actions performed during execution. The authority granted by a user may be expressed in a structured policy, carried by a credential, and checked when a tool invokes a service. The approved limits must remain enforced throughout this process, even when components describe resources and operations differently.

Correct credential validation can still allow actions that the approved policy forbids. In Figure~\ref{fig:rethink}(\subref{fig:representation}), the user's grant permits sending summary $S$ to Alice, and the structured policy correctly records the permitted content and recipient. When issuing a credential, however, the system reduces these restrictions to a generic permission to send mail. If the mail service receives only this permission, it can validate the credential correctly while lacking the recipient and content restrictions needed to enforce the approved policy. Verifying the credential's authenticity does not recover these missing restrictions. They must be included in the service's checks or enforced before the request reaches it.

Expressing the user's authorization correctly and preserving it during execution are separate obligations. The first concerns whether the policy captures what the user permits; the second concerns whether the implementation enforces that policy. In the mail example, the policy is correct, but its translation loses the recipient and content restrictions. When asking the user to approve a policy, the system should clearly show these limits and record the approved version, so that subsequent checks enforce the same policy.

Enforcing a policy requires knowing which actual resources and service operations its restrictions refer to. AC4A~\cite{ac4a} relies on application-provided functions that identify the resource permissions sufficient for an API call, based on the requested operation and its arguments. ceLLMate~\cite{cellmate} links high-level actions, such as making a purchase, to the HTTP requests that carry them out. Fides~\cite{fides} uses trusted code around tool calls to assign or propagate labels describing the integrity and confidentiality of their results. These connections determine how policy restrictions apply to actual calls and data. Checking a rule correctly is insufficient if the check uses an incorrect description of the resources accessed, the action performed, or the data returned.

Every resource access or external action allowed by the implementation must also be permitted by the approved policy. Let $P$ be an approved policy, $M(P)$ the credentials and enforcement rules used to implement it, $\mathcal{E}(M(P))$ the effects allowed by those controls, and $\alpha$ a mapping that expresses those effects in the terms used by the policy. Preserving the restrictions requires
\begin{equation}
\alpha\bigl(\mathcal{E}(M(P))\bigr)\subseteq\llbracket P\rrbracket,
\label{eq:refinement}
\end{equation}
where $\llbracket P\rrbracket$ denotes the set of abstract effect traces allowed by the approved policy, such as permitted operations paired with resolved resource, recipient, and content attributes, and the abstraction mapping $\alpha$ lifts concrete runtime invocations and arguments in $\mathcal{E}(M(P))$ to that policy vocabulary. In the mail example, allowing delivery to Bob produces a concrete transmission whose image under $\alpha$ falls outside $\llbracket P\rrbracket$, violating this condition even if the credential itself is valid. An implementation may block some operations that the policy permits, preserving its restrictions at the cost of useful work, whereas allowing an operation outside $\llbracket P\rrbracket$ requires additional authorization. Equation~\ref{eq:refinement} can be checked mechanically when both $P$ and the tool's effect descriptions are expressed in a decidable fragment, such as Datalog or finite-domain predicates, $\alpha$ soundly over-approximates concrete argument-to-resource resolution, and execution confinement prevents unmediated side channels. For ordering or cumulative policies, $\llbracket P\rrbracket$ and $\alpha$ range over execution histories.

One design direction is to use \emph{effect contracts}: explicit descriptions of the resources a tool operation may access, the data it may send, the recipients of those data, and the changes it may make to external systems. These contracts would allow an enforcement component to compare the possible effects of a tool call with the approved policy before permitting execution. If a contract lists effects that the tool cannot actually produce, checks based on that contract may block permitted work. If it omits possible effects, those checks may miss unauthorized behavior. The descriptions must also be updated when tool behavior changes. A tool that begins forwarding data to an external service introduces a new disclosure, even if its name and accepted arguments remain unchanged. The system therefore needs to reassess whether the existing grant permits the tool's changed behavior.

\subsection{Authority Has State beyond Individual Credentials}\label{sec:rethink-state}
A valid credential does not by itself establish whether an action is still authorized: another agent may have already spent the full amount authorized for the task, an earlier attempt to perform the same action may already have completed, or the grant may have been revoked. Authorization therefore depends on what has happened during execution, and on permissions originally granted.

\begin{figure}[t]
\centering
\Description{Parallel purchasing agents share a one-item task allowance. A timeline separates grant changes, request submissions, and uncertain remote effects. Time proceeds downward; revocation blocks a retry without resolving the original operation's outcome.}
\begin{subfigure}[t]{.43\linewidth}
\centering\includegraphics[width=.92\linewidth]{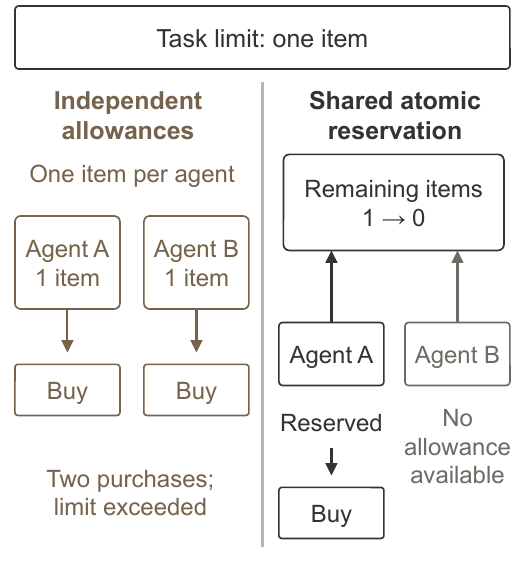}
\caption{Coordinating a shared purchase limit}\label{fig:shared}
\end{subfigure}
\begin{subfigure}[t]{.43\linewidth}
\centering\includegraphics[width=.92\linewidth]{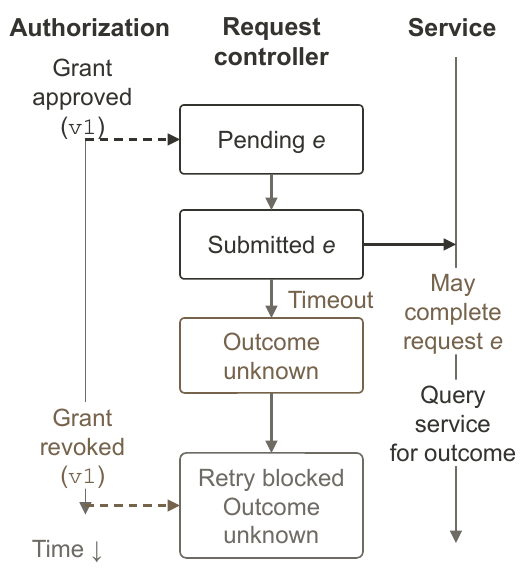}
\caption{Revocation with uncertain outcomes}\label{fig:temporal}
\end{subfigure}
\caption{Authorization depends on shared limits and execution outcomes. (a) Separate per-agent allowances can permit two purchases despite a one-item task limit; a shared reservation prevents this excess. (b) Revocation blocks a retry but leaves the original request's outcome unresolved. In (b), time runs downward; v1 denotes the grant version, and $e$ identifies the same operation across attempts.}
\label{fig:state}
\end{figure}

Delegation and retries can both reuse permission intended for a single purchase. Figure~\ref{fig:state}(\subref{fig:shared}) illustrates an agent authorized to purchase one item. It delegates to two purchasing agents, each permitted to buy one item. Each purchasing agent stays within its own limit, yet together they can buy two items. These individual limits do not preserve the task's overall limit. Retries create a similar risk across attempts: after a timeout, the original purchase may already have completed, so repeating it could buy a second item. Checking each agent or attempt misses the combined limit.

Existing mechanisms track changes that affect whether granted permissions remain usable. SAGA checks token validity and limits requests, MiniScope invalidates session tokens that depend on a revoked user authorization, and ceLLMate applies policies that account for execution state~\cite{saga,miniscope,cellmate}. Combining these mechanisms requires consistent records of the grant, how much of the task's limit remains available, and requests whose outcomes are still unknown. A credential can identify the grant or refer to these records, but checking it in isolation does not establish what the workflow is still permitted to do.

Coordinating a shared limit requires recording which operations have reserved or used part of it. Before a purchase, the system can reserve part of the permitted quantity for that operation, making it unavailable to other agents. Checking the remaining quantity and reserving it must happen as one indivisible step, so two agents cannot both reserve the last item. In Figure~\ref{fig:state}(\subref{fig:shared}), Agent A reserves the right to buy the only permitted item, leaving none for Agent B. Retries also need an identifier that stays the same across attempts to perform the same operation. A service that supports duplicate detection can use this identifier to recognize a repeated request and avoid executing the operation twice. Recording the grant version that authorized the operation, and checking whether that grant has been revoked, connects these execution records to the applicable authorization. Together, these checks account for permission changes and prior use.

While atomic reservation, idempotency keys, and compensating Sagas~\cite{sagas} are well established in distributed systems, they do not by themselves settle agent authorization state. Classical transaction mechanisms assume deterministic retries with fixed identifiers and reversible state transitions, whereas an agent reacting to a timeout may re-plan semantically, altering arguments, splitting a request, or switching tools, so that a repeated attempt no longer shares the syntactic identifier $e$. Moreover, agent workflows introduce revocation across dynamic delegation graphs, time-bounded human approvals, and external disclosures that compensating actions cannot undo.

What revocation can prevent depends on how far an operation has progressed. Revoking access to a data source can block future reads, while information already stored by the workflow requires controls on its subsequent use and release. Once a request has been submitted, preventing its effect requires coordination with the service executing it; revoking the requesting agent's credential alone does not cancel that request. Information already delivered to a recipient cannot be recalled through access revocation. Revocation semantics therefore need to depend on the operation's current status.

An unknown outcome affects both whether a request may be retried and whether the quantity reserved for its operation can be used elsewhere. Figure~\ref{fig:state}(\subref{fig:temporal}) shows a request whose outcome remains unknown after a timeout. Revoking its grant blocks a retry but does not establish whether the service completed the original request. A service may prevent duplicate execution when a retry is authorized without reporting whether the original request completed. Resolving that uncertainty requires querying the service about the original request or obtaining a record that confirms its outcome. Until then, the workflow should retain a record of the request, its unknown outcome, and the authorization under which it was submitted. For a purchase, this means keeping track of an item that may already have been bought, rather than assuming it remains available within the task's limit. Later authorization decisions can account for effects that may already have occurred.

\section{Open Challenges and Future Directions}\label{sec:challenges}
The preceding analysis reframes how authorization should be understood and maintained in agent workflows. Translating these insights into practical systems presents challenges in system design, user interaction, and validation. Table~\ref{tab:agenda} summarizes these challenges and potential directions.

\begin{table}[!htbp]
\caption{Open challenges and potential research directions for agent access control.}
\label{tab:agenda}
\begingroup
\fontsize{9}{11}\selectfont
\setlength{\tabcolsep}{4pt}
\renewcommand{\arraystretch}{1.12}
\begin{tabularx}{\linewidth}{@{}P{.21\linewidth}YY@{}}
\toprule
Research Area & Open Challenge & Potential Direction \\
\midrule
Planning and approval & Find feasible plans with limited access and user effort & Authorize exploration first; reuse approvals within explicit limits \\
\addlinespace[4pt]
Context handling & Preserve source roles and approval limits through summaries and delegation & Link generated claims to supporting evidence and source roles \\
\addlinespace[4pt]
Tool effects & Track indirect effects and\newline changes in tool behavior & Combine program analysis, execution records, and comparisons across versions \\
\addlinespace[4pt]
Recovery and revocation & Coordinate pending actions across services with different capabilities & Base recovery on cancellation, duplicate prevention, and outcome reporting \\
\addlinespace[4pt]
Benchmark construction & Check authorization across alternative plans and changing grants & Check actions against predefined grants under controlled workflow changes \\
\bottomrule
\end{tabularx}
\endgroup
\end{table}

\subsection{Planning with Limited Authority and User Intervention}
One challenge in agent planning is \emph{allowing enough exploration to identify feasible plans without granting unnecessary access}. The agent may need to inspect resources before it can determine which permissions different plans require. Yet gathering enough information to choose a plan may require less access than executing it. This suggests staged authorization: explicitly authorize limited exploration, use what the agent learns to seek an adequate plan within the current grant, and present any proposed expansion as a separate approval request.

A second challenge is \emph{reducing repeated approval requests while preserving authorization boundaries}. Several tool calls may implement a single user-approved decision, so a new call need not require a new approval if it remains within that decision's scope. This motivates reusing approvals while their conditions remain valid and grouping related operations into a clearly scoped request. User involvement can then focus on proposed permission expansions and ambiguities that affect whether an action is authorized. Approval interfaces should explain what additional access a request would permit and whether less intrusive alternatives exist. These approaches should reduce approval frequency and user effort while preserving task quality and authorization boundaries.

\subsection{Preserving Source Roles across Context Transformations}
When agents summarize information or pass it to other agents, a key challenge is \emph{preserving source roles and approval limits}. A summary may combine factual observations with an approval, obscuring which source supports each statement and what was actually approved. The original evidence can remain available even when its wording changes. This suggests linking generated claims to that evidence and recording which decisions each source may influence.

These links could support checks of both factual meaning and authorization scope. For each claim used to justify an action, the system could check whether the original statement supports it and whether the source is authorized to supply that evidence or approve the action. For example, an approval to send a report to one recipient should retain that limit when summarized. The research difficulty is maintaining these links when claims from several sources are combined or rewritten. These approaches should preserve useful contextual information while keeping subsequent decisions consistent with source roles and original approval limits.

\subsection{Maintaining Effect Contracts across Tools and Services}
For tool access control, a key challenge is \emph{describing effects that are not visible in the tool's interface}. A call may access additional resources, send data to another service, or schedule an action that occurs later. These effects are difficult to capture when the service's implementation cannot be inspected. Different sources offer complementary information: program analysis can identify possible operations, execution records show what occurred, and developer descriptions can explain behavior inside external services. When external services cannot be inspected statically, combining interface schemas with sandboxed dry runs, outbound data-flow tracking, and runtime checks on returned state can surface indirect side effects. Integrating these sources into effect contracts captures resource accesses, data transfers, and external actions while marking uncertain effects.

Keeping these descriptions useful requires \emph{updating them as tool behavior changes without extensive manual review}. A tool can introduce a new recipient or external action without changing its interface, motivating comparisons of effect descriptions across versions. One direction is to compare code changes, developer specifications, and observed network behavior across service updates to detect silent shifts in tool behavior, flagging newly introduced disclosures or state changes for review. Such reviews should determine whether existing grants cover the changed behavior or additional approval is needed.

\subsection{Recovery and Revocation across Service Boundaries}
When an agent relies on external services, a key challenge is \emph{determining what recovery and revocation can guarantee across different service capabilities}. Some services can cancel submitted requests, prevent duplicate execution, or report whether a request completed; others provide only part of this support. Each capability addresses a different need: cancellation may prevent a pending action, duplicate detection can make an authorized retry safe from repeated effects, and status reporting can resolve an unknown outcome. These differences suggest deriving recovery rules from the capabilities each service provides. Research could establish which guarantees hold across capability combinations when multiple services and delegated agents participate.

An unresolved request also creates the challenge of \emph{coordinating shared limits while its outcome remains unknown}. After a timeout, a purchase may already have completed. Releasing its reserved allowance could permit an extra purchase, while holding it indefinitely could block useful work. Because one agent's unresolved request affects what others may do, recovery needs records shared across the workflow. A potential direction is to link pending requests, their grants, and reserved allowances, then update these records as services report completion or cancellation. Where an effect has already occurred, a compensating action~\cite{sagas} such as a refund may support recovery, but cannot reverse an unauthorized disclosure or prevent semantic retries across alternative tools.

\subsection{Building Benchmarks for Agent Access Control}
Building a benchmark for agent access control involves \emph{defining correct authorization without prescribing one execution plan}. Different plans can complete the same task, and permission to perform an action can change during execution. AgentDojo's stateful tasks and separate utility and attack-outcome measurements provide a starting point~\cite{agentdojo}. Task-level restrictions can describe permitted accesses and effects across different plans. This suggests pairing each task with explicit grants and independently specified rules for checking recorded effects. These rules account for grant changes and request states, allowing multiple authorized plans to satisfy the task.

A further challenge is \emph{covering unused permissions and failures absent from successful runs}. A normal run can leave excessive access unused and avoid the timeouts or concurrent requests that expose coordination failures. This motivates controlled workflow environments that record resource accesses, data transfers, and external actions while injecting changes in information sources, delegation, and service responses, including over-privileged plan choices, context-induced grant expansion, credential translation loss, and concurrent retry races under revocation. Attempts outside the grant measure \emph{authorization correctness} by separating blocked requests from executed violations and permitted plans that are falsely rejected. Comparing alternative plans and unused permissions measures \emph{enforcement completeness}, capturing both excess authority exposure and unauthorized effects that remain enabled. Recording user approval interactions and execution cost further captures \emph{usability}, such as task completion and interruption frequency, alongside \emph{runtime overhead} in policy latency and token consumption. Such environments could support comparable studies of protection and cost under shared task requirements and dynamic grant changes.

\section{Related Work}
\emph{Security foundations.} Least privilege and complete mediation~\cite{saltzer}, privilege separation~\cite{privsep}, and capability-based sandboxing~\cite{capsicum} establish ways to limit authority and isolate execution. Models of information flow~\cite{denning} govern data propagation, while enforceable-policy theory~\cite{schneider} characterizes runtime monitoring guarantees. Together, these foundations connect permission assignment with control over actions and information.

\noindent\emph{Agent authorization and user control.} Existing surveys and vision papers examine complementary slices of the problem: Michael and Roesner~\cite{permissions} focus on user-facing permission interfaces and interaction timing, Surapani et al.~\cite{architectures} categorize deployment architectures for policy decision and enforcement points around tool calls, and Li et al.~\cite{aac} outline a vision adapting classical access-control models to contextual information release. Whereas those works organize the space by user interaction, architectural placement, or classical access models, we compare mechanisms across three orthogonal dimensions, namely control objects, enforcement boundaries, and underlying dependency assumptions, to analyze how authorization guarantees compose across an evolving workflow. This cross-boundary view directly motivates our four rethinking perspectives: integrating authority exposure into plan selection alongside policy-aware planning~\cite{prudentia,elasticprivilege}, tying contextual influence to source decision roles beyond user preference elicitation~\cite{dataperms,janus,userpolicies}, preserving restriction semantics across policy-to-credential translation, and maintaining shared authorization state across delegation and uncertain outcomes.

\noindent\emph{Agent security evaluation.} AgentDojo~\cite{agentdojo} evaluates prompt-injection attacks and defenses in stateful tasks, separating task utility from attack outcomes. ToolEmu~\cite{toolemu} emulates tool execution with language models to identify potential failures. ToolPrivBench~\cite{toolprivbench} evaluates whether agents choose higher-privilege tools despite sufficient lower-privilege alternatives. These benchmarks provide foundations for evaluating authorization across alternative plans, changing grants, and external effects.

\noindent\emph{Requirements and software ecosystems.} Harel and Marron~\cite{requirements} examine the difficulty of discovering and describing intended behavior, beyond implementing a given specification. This perspective helps separate two authorization questions: whether a policy expresses the user's intended delegation and whether execution respects that policy. Zhao et al.~\cite{appstore} study LLM app stores through stakeholder relationships, security risks, and governance needs. Their ecosystem perspective is relevant when agents combine services supplied by different developers and platforms. Together, these perspectives connect the content of authorization requirements with responsibility for defining and enforcing them. They motivate explicit treatment of whose decisions a policy represents and where those decisions take effect across service boundaries.

\section{Conclusion}
Agent access control should enable useful work while keeping execution within delegated authority. We surveyed recent approaches to policy specification, identity and delegation, action enforcement, and information-flow control. Our analysis highlights the need to preserve authorization as agents choose plans with different permission requirements, interpret context from sources with distinct authority, translate approved policies into operations, and respond to changes during execution. Further progress requires connecting these mechanisms across workflow boundaries and building benchmarks that capture changing grants and execution conditions. The aim is to preserve approved limits throughout a task while reducing the burden of permission management on users.

\label{main-text-end}
\bibliographystyle{ACM-Reference-Format}
\bibliography{references}
\end{document}